\documentclass[pra,amsmath,amssymb,twocolumn,superscriptaddress]{revtex4}

\usepackage{amsmath,amssymb}
\usepackage[usenames]{color}
\usepackage{mathbbol}
\usepackage{amssymb}
\usepackage{grffile}
\usepackage[pdftex]{graphicx}
\usepackage{amsmath, amstext, amssymb, amsfonts, amsxtra}
\usepackage{textcomp}
\usepackage{xspace}
\DeclareSymbolFontAlphabet{\amsmathbb}{AMSb}

\newcommand{\aop}{\hat{a}}

\newcommand{\nop}{\hat{n}} 
\newcommand{\nbar}{\bar{n}}     
   
\newcommand{\adop}{\hat{a}^{\dagger}}

\newcommand{\cur}{\mathcal{J}}

\newcommand{\rhop}{\hat{\rho}}

\newcommand{\Hop}{\hat{H}} 
\newcommand{\Dop}{\mathcal{D}}
\newcommand{\lan}{\left\langle} 
\newcommand{\ran}{\right\rangle} 
 
\newcommand{\im}{{\rm i}}

\newcommand{\tun}{\mathcal{C}}       
 
\newcommand{\Lop}{\mathcal{L}}

\newcommand{\sutd}{EPD Pillar, Singapore University of Technology and Design, 8 Somapah Road, 487372 Singapore} 

\begin{document}

\title{Dissipatively driven strongly interacting bosons in a gauge field }                     

\author{Chu Guo} 
\affiliation{\sutd}
\author{Dario Poletti}
\affiliation{\sutd}

\begin{abstract} 
The interplay between dissipation, interactions and gauge fields opens the possibility to rich emerging physics. Here we focus on a set-up in which the system is coupled at its extremities to two different baths which impose a current. We then study the system's response to a gauge field depending on the filling. 
We show that while the current induced by the baths has a marked dependence on the magnetic field at low fillings which is significantly reduced close to half-filling. We explain the interplay between interactions, gauge field and dissipation by studying the system's energy spectrum at the different fillings. This interplay also results in the emergence of negative differential conductivity. For this study we have developed a number-conserving treatment which allows a numerical exact treatment of fairly large system sizes, and which can be extended to a large class of systems.             
\end{abstract}


\maketitle

A deeper understanding of the far-from equilibrium transport properties of complex quantum systems would lead to fascinating progress for future nanotechnologies. A particularly interesting challenge is that of characterizing and controlling the transport properties of quantum systems. In such systems manybody effects  can induce, remove or shift phase transitions lines. This significantly affects the properties of a system and our ability to control it. 
Another salient tool used to control transport and induce new phases of matter is a gauge field. The quantum Hall effect is a paramount example of the role of gauge fields on transport properties \cite{KliltzingPepper1980, Laughlin1981}. In a type-$II$ superconductor the increase of the magnitude of the magnetic field can drive a transition from a diamagnetic Meissner phase to a superconductor with an Abrikosov vortex lattice \cite{Abrikosov1957}. Adding interactions to this system can lead to even more exotic phases of matter with topological order \cite{TsuiGossard1982, Laughlin1983, ChenWen2010}.     
  
\begin{figure}
\includegraphics[width=0.8\columnwidth]{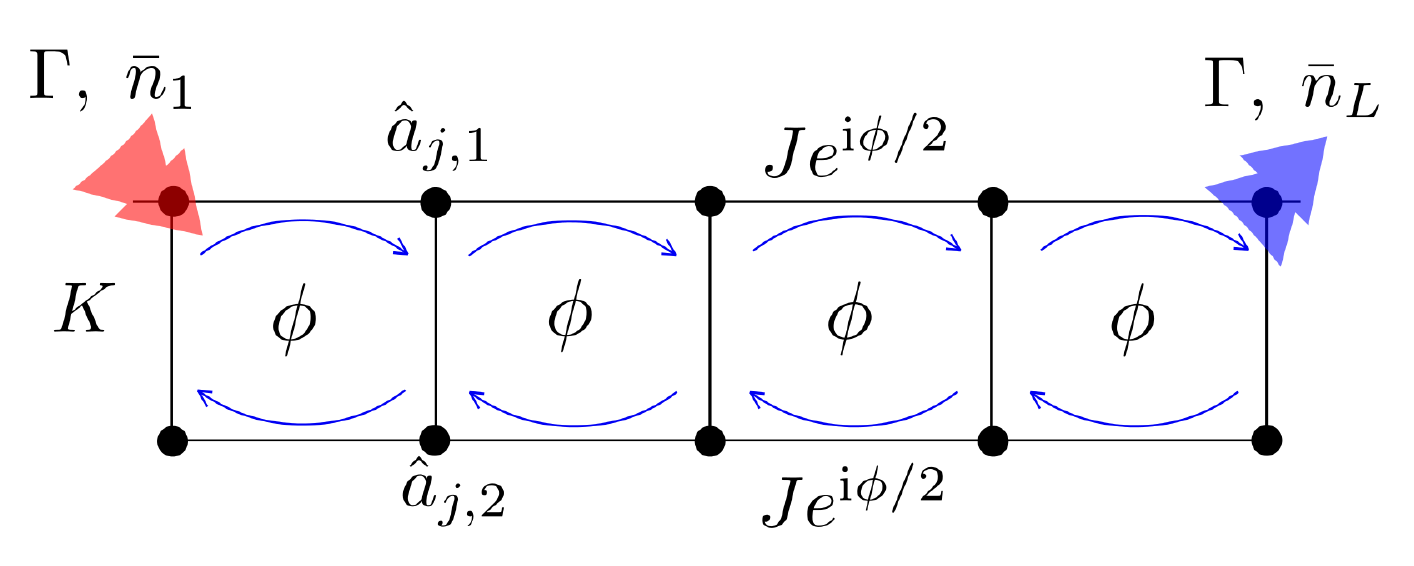}
\caption{(color online) Ladder made of two coupled linear chains, referred to as legs, with local bosonic excitations described by the annihilation operators at site $j$, $\aop_{j,p}$, where $p=1,2$ respectively for the upper and the lower leg. $K$ is the tunnelling amplitude between the legs, on what are referred to as rungs of the ladder, while $J$ is the amplitude of tunnelling between sites in the legs. A gauge field imposes a phase $\phi$. The coupling to the baths is represented by the thick arrows. Each bath is characterized by the average density of bosons $\bar{n}_j$ the bath itself imposes on the rung $j$ and the strength of the coupling $\Gamma$. 	
} \label{fig:fig1} 
\end{figure}

The minimal set-up in which such rich phenomenology can be explored is that of coupled chains (a ladder) with a gauge field as the one depicted in Fig.\ref{fig:fig1}. This system has attracted intense theoretical scrutiny \cite{Kardar1986, Granato1990, DennistonTang1995, Nishiyama2000, OrignacGiamarchi2001, ChaShin2011, DharParamekanti2012, DharParamekanti2013, CrepinSimon2011, TovmasyanHuber2013, PetrescuLeHur2013, WeiMueller2014, TokunoGeorges2014, PiraudSchollwock2015, GreschnerVekua2015,  GreschnerVekua2016, CavalneseStrinatiMazza} and it has been experimentally studied both with Josephson-junctions arrays \cite{ZantMoiij1992, OudenaardenMooij1996, OudenaardenMooij1998, FazioZant2001} and with ultracold gases with bosons and fermions \cite{AtalaBloch2014, ManciniFallani2015, TaiGreiner2016}, thanks to the use of synthetic gauge fields \cite{DalibardOhberg2011, GoldmanSpielman2014}. 

Here we will consider a quantum system in presence of a gauge field and connected at its extremities to two different baths which would impose a current through it (so called boundary dissipatively driven systems). Ion trap experiments promise to be an ideal set-up for a clean realization and study of such transport problems \cite{BermudezPlenio2013, RammHaffner2014, GuoPoletti2015}. The set-up could be realized using ion micro-cavity arrays with a nonlinear local potential \cite{PorrasCirac2004, DengCirac2008, DuttaSengupta2013, DuttaSengupta2016, GuoPoletti2015}, for which the gauge field is generated by Raman coupling \cite{BermudezPorras2011}.  It should be noted that magnetic fields have already been used to effectively modify heat transport in Josephson junctions \cite{GiazottoMartinezPerez2012}. 

Recently, a boundary driven coupled chains of free bosons under the effect of a gauge field was studied \cite{GuoPoletti2016}. There it was shown that, depending on how the baths where coupled to the ladder, the chiral current changes abruptly on two phase transition lines, implying the emergence of a previously unpredicted non-equilibrium phase transition. At this transition line, coinciding with the opening of a gap in the bulk spectrum, the total current through the system also starts to change significantly, hence the gauge field can be used to strongly control the current flow.      

However it is necessary to gain a deeper understanding of how the interplay between the gauge fields and interactions between the bosons will affect the transport. Hence in this paper we are going to study the steady state transport properties of hardcore bosons driven out of equilibrium by dissipative boundary driving. We will show that the controllability of the current via a gauge field (i.e. the ability to alter the current), is significantly reduced as the average filling is increased from low to near half-filling. We will also show that the non-linear dependence of the current on the density also results, depending on the gauge field, in the emergence of negative differential conductance. 

It is important to stress that, in order to analyze this system, we introduce an exact numerical approach with conserved quantum numbers to compute the steady state. With this method we are able to readily study the exact density matrix of a system with up to $14$ sites. Moreover this method also allows us to gain a much deeper insight on the working of the system.

{\it Model:} We study a ladder made of two coupled chains (or legs) each composed of $L$ sites. Two sites in different chains which are coupled form a rung. The ladder is coupled at its extremities to two baths which can inject or remove bosons at different rates. The set-up is depicted in Fig.\ref{fig:fig1}. The evolution of the density operator $\rhop$ is given by a master equation with a Lindbladian $\Lop$ 
\begin{equation}
\frac{d\rhop}{dt} = \mathcal{L}(\rhop) = - \frac{\im}{\hbar}\left[\Hop,\rhop\right]+\Dop(\rhop)\label{eq:MME}
\end{equation}   
where the Hamiltonian $\Hop$ is given by 
\begin{align}
\Hop=-&J \sum_{p,j} e^{\im\Phi_p}\adop_{j,p}\aop_{j+1,p} - K\sum_{j} \adop_{j,1}\aop_{j,2}  + {\rm H.c.} \label{eq:Ham} 
\end{align}
Here $K$ is the tunnelling constant in the rungs, $J$ for the legs, and the phase of the tunneling in the legs $\Phi_p=(-1)^{(p-1)}\phi/2$ such that a particle doing a loop around one plaquette acquires a phase $\phi$.  
The operators $\aop_{j,p}$ ($\adop_{j,p}$) annihilates (creates) a boson in the upper ($p=1$) or lower ($p=2$) chain at the $j$-th rung of the ladder. As we focus on the role of strong interactions, we consider hardcore bosons, for which at most one boson can occupy one site, i.e. $\adop_{j,p}\adop_{j,p}=0$. The dissipator in Lindblad form \cite{Lindblad1976, GoriniSudarshan1976} is given by     
\begin{align} 
\Dop(\rhop)  = \sum_{j=1,L} \Gamma & \allowbreak\left[ \allowbreak(1-\bar{n}_{j}) \left(\aop_{j,1}\rhop\adop_{j,1} - \aop_{j,1}\adop_{j,1}\rhop \right) \right. \nonumber \\
\allowbreak &+  \left. \bar{n}_{j}\right. \allowbreak  \left.\left(\adop_{j,1}\rhop\aop_{j,1} - \adop_{j,1}\aop_{j,1}\rhop \right) \allowbreak + {\rm H.c.} \right] \label{eq:diss}   
\end{align}
where $j=1$ or $L$, $\Gamma$ is the overall coupling constant. The dissipator tends to set the local density at site $(j,1)$ to the value $\bar{n}_{j}$ if decoupled from the others \cite{gamma}. The baths will thus induce a particle current when $\Delta\bar{n}=\bar{n}_{1} - \bar{n}_L \ne 0$.

{\it Total current}: We focus our attention on the particle current at steady state \cite{unique}. The total current through the system, $\cur$, is given by the sum of the current in each leg $\cur=\sum_p\cur^L_{j,p}$ where the current in the $p$-leg is  
$\cur^L_{j,p} = \lan\im Je^{\im(-1)^{p+1}\phi/2}\aop_{j,p}^{\dagger}\aop^{}_{j+1,p} +{\rm H.c}\ran /\hbar$ and $\lan\dots\ran$ means the expectation value for the steady state. The current in the rungs is instead given by $\cur^R_{j,1\rightarrow 2} = \lan \im K \aop_{j,1}^{\dagger}\aop_{j,2} + {\rm H.c.} \ran /\hbar$. 
The leg and rung currents are associated to the continuity equations $\frac{\partial \lan\nop_{j,1}\ran}{\partial t} = \cur^L_{j-1,1} - \cur^L_{j,1} - \cur^R_{j,1\rightarrow 2}$ and $\frac{\partial \lan\nop_{j,2}\ran}{\partial t} = \cur^L_{j-1,2} - \cur^L_{j,2} + \cur^R_{j,1\rightarrow 2}$ for $1<j<l$. 

\begin{figure}
\includegraphics[width=\columnwidth]{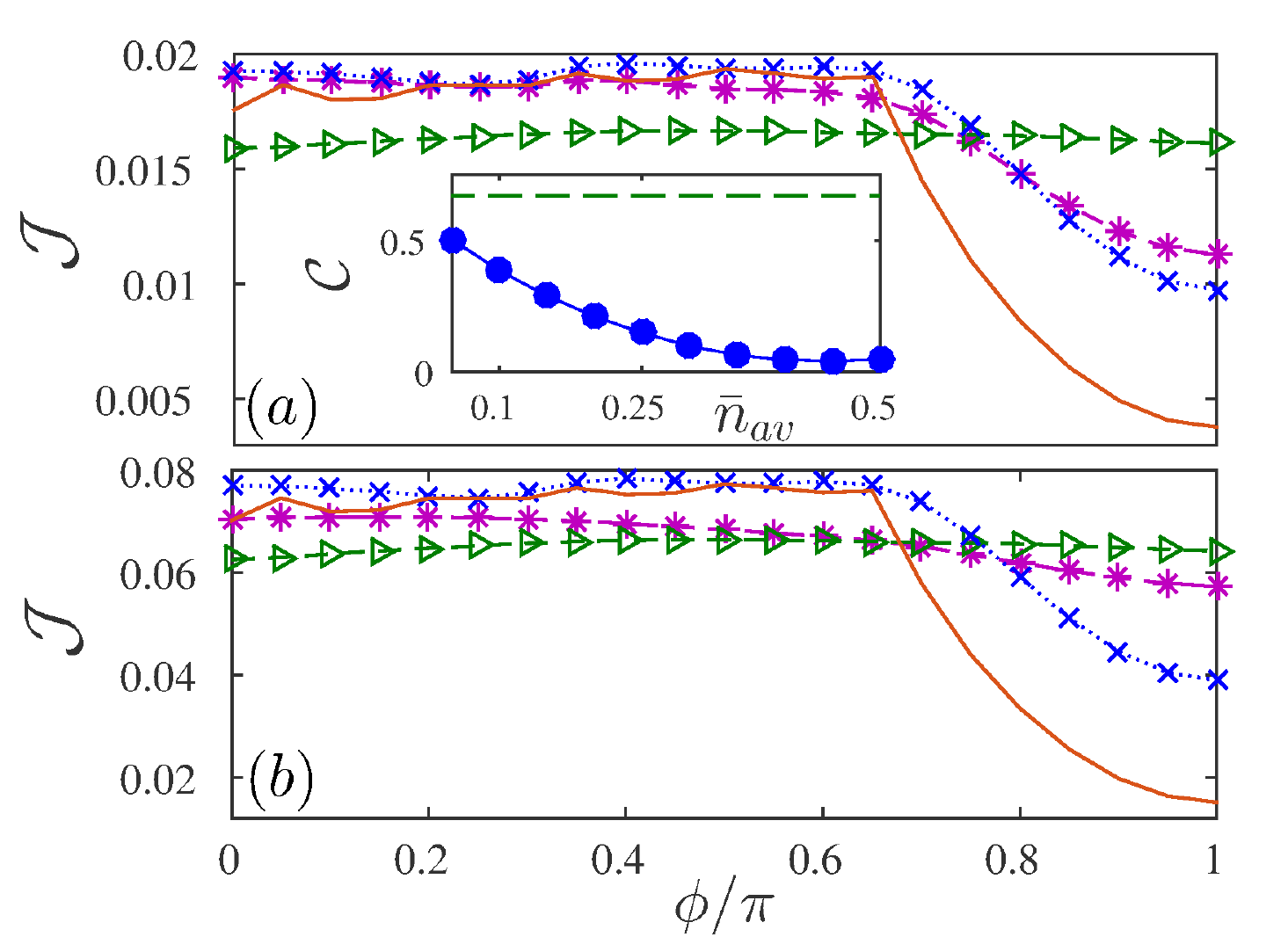} 
\caption{(color online) Total current $\cur$ versus phase $\phi$ for different values of average density $\nbar$. As the system approaches $\nbar_{av}=0.5$ the current becomes much less sensitive to the phase. (a,b) The continuous red lines are for free bosons and a ladder of length $L=200$. The other three lines are for $L=6$ and free bosons (blue dotted line with $\times$), $\bar{n}_{av}=0.1$ (purple dashed line with $*$) and $\bar{n}_{av}=0.5$ (green dashed line with triangles). In panel (a) $\Delta \bar{n}=0.1$, while in panel (b) $\Delta \bar{n}=0.4$. The inset of panel (a) shows the controllability $\tun$ versus $\bar{n}_{av}$ for $\Delta \bar{n}=0.1$ (circles) and for free bosons (dashed green line). The controllability is strongly reduced as the average filling increases.} \label{fig:fig2} 
\end{figure}

In Fig.\ref{fig:fig2}(a) we show the total current in the system as a function of the gauge field $\phi$ for a small value of $\Delta\bar{n}$ in panel (a) ($\Delta\bar{n}=0.1$) and a larger value in panel (b) ($\Delta\bar{n}=0.4$). It is important to have as a reference the case of non-interacting bosons which shows a marked dependence on the phase. The red-continuous thin line shows the current for a large ladder ($L=200$) of non-interacting bosons. This was computed in \cite{GuoPoletti2016} and it shows the significant change in the total current at the critical gauge field $\phi_c=2\pi/3$. However, because of the computational complexity of a ladder of hardcore bosons, in this work we are limited to short ladders, which, as shown later, already manifest remarkable effects. In order to fairly compare the strongly interacting bosons to the non-interacting ones, we show, with the blue-dotted line with crosses, the total current versus $\phi$ for a ladder of $6$ rungs of free bosons. The curve is smooth and has larger oscillations, however it still shows a marked dependence on the phase $\phi$, and it shows a strong current suppression for $\phi$ approaching $\pi$. This means that even for a small chain, the gauge field can be used to control the current.    
We now consider the case of hardcore bosons, from low average filling, to around half-fillings. At low average density bias $\bar{n}_{av}=(\bar{n}_1+\bar{n}_L)/2$, the behavior of the current as a function of the phase naturally resembles that of non-interacting bosons (see pink dashed line with $*$). However, as we increase $\bar{n}_{av}$ such that the local occupation reaches half-filling, the difference between the response of the free bosons compared to the hardcore bosons is striking. In particular, close to half-filling the dependence of the current on the gauge field is highly reduced, as shown by the bold green continuous line.  
One direct consequence is that, at larger fillings, the current can be much larger than for free bosons because the gauge field cannot significantly reduce it (see the region for $\phi$ close to $\pi$).   
This may seem surprising from an analysis of the groundstate because, near half-filling, the spectrum of the hardcore boson ladder is gapped. In the following we will explain the mechanisms behind this behavior.  
Fig.\ref{fig:fig2}(b) shows a similar behavior but with less marked difference because in this panel $\Delta \bar{n}=0.4$.  
In the inset of Fig.\ref{fig:fig2}(a) we show the controllability $\tun$, which describes how well the gauge field can tune the current in the system. For given dissipative boundary driving and tunneling parameters, the controllability is given by $\tun=\left[\max_\phi \cur -\min_{\phi}\cur\right]/\left[\left(\max_{\phi}\cur+\min_\phi(\cur)\right)/2\right]$. The inset shows that, for $\Delta \bar{n}=0.1$, the controllability $\tun$ is suppressed by one order of magnitude as the average filling increases. Moreover, even at low fillings the controllability is lesser than that for free fermions (green dashed line).

%

%
%
%


{\it Exact numerical approach with quantum numbers}: To gather a deeper understanding and to be able to analyze in a numerically exact way this system, we have developed an approach which takes into account the total quantum number. We first write the density operator as 
 
\begin{align}
\rhop&=\sum_{\vec{m}_N,\vec{m}'_{N'},N,N'} \rho^{\vec{m}_N,N}_{\vec{m}'_{N'},N'} |\vec{m}_N,N\rangle\langle\vec{m}'_{N'},N'|     
\end{align}
where $N$ ($N'$) are the total number of particles respectively for the ket (bra), while $\vec{m}_N$ ($\vec{m'}_{N'}$) described the distribution of the $N$ ($N'$) atoms between the $2L$ sites. It should be noted that the Hamiltonian $\Hop$ in Eq.(\ref{eq:Ham}) conserves the total number of atoms either in the bra or in the ket. Moreover the dissipator $\Dop$ in Eq.(\ref{eq:diss}) only couples an element $\rho^{\vec{m}_N,N}_{\vec{m}'_{N'},N'}$ with another element $\rho^{\vec{n}_{N\pm 1},N \pm 1}_{\vec{n}'_{N'\pm 1},N'\pm 1}$ where the total number of particles in the ket and bra is either increased or decreased by one particle. Last it should be pointed out that since the steady state is unique and the initial condition can be chosen to be in a pure state, it is possible to derive that the steady state will be exactly described by a much simpler ansatz of the form 
\begin{align}
\rhop_{ss}&=\sum_N \rhop_{ss}^N = \!\! \sum_{\vec{m}_N,\vec{m}'_{N},N} \!\!     \rho^{\vec{m}_N,N}_{\vec{m}'_{N},N} |\vec{m}_N,N\rangle\langle\vec{m}'_{N},N|   \label{eq:rhoss}     
\end{align}
where the total number of particles in the bra or in the ket is the same. Hence, to find the steady state we can,  for instance, use a state within one number block as an initial condition and evolve it using the ansatz in Eq.(\ref{eq:rhoss}) with the master equation (\ref{eq:MME}). For chains up to $7$ rungs ($14$ sites), we compute the steady state of the Lindbladian $\mathcal{L}$ by directly solving the linear equation $\mathcal{L}(\rhop)=0$ with Arpack. 

\begin{figure}
\includegraphics[width=\columnwidth]{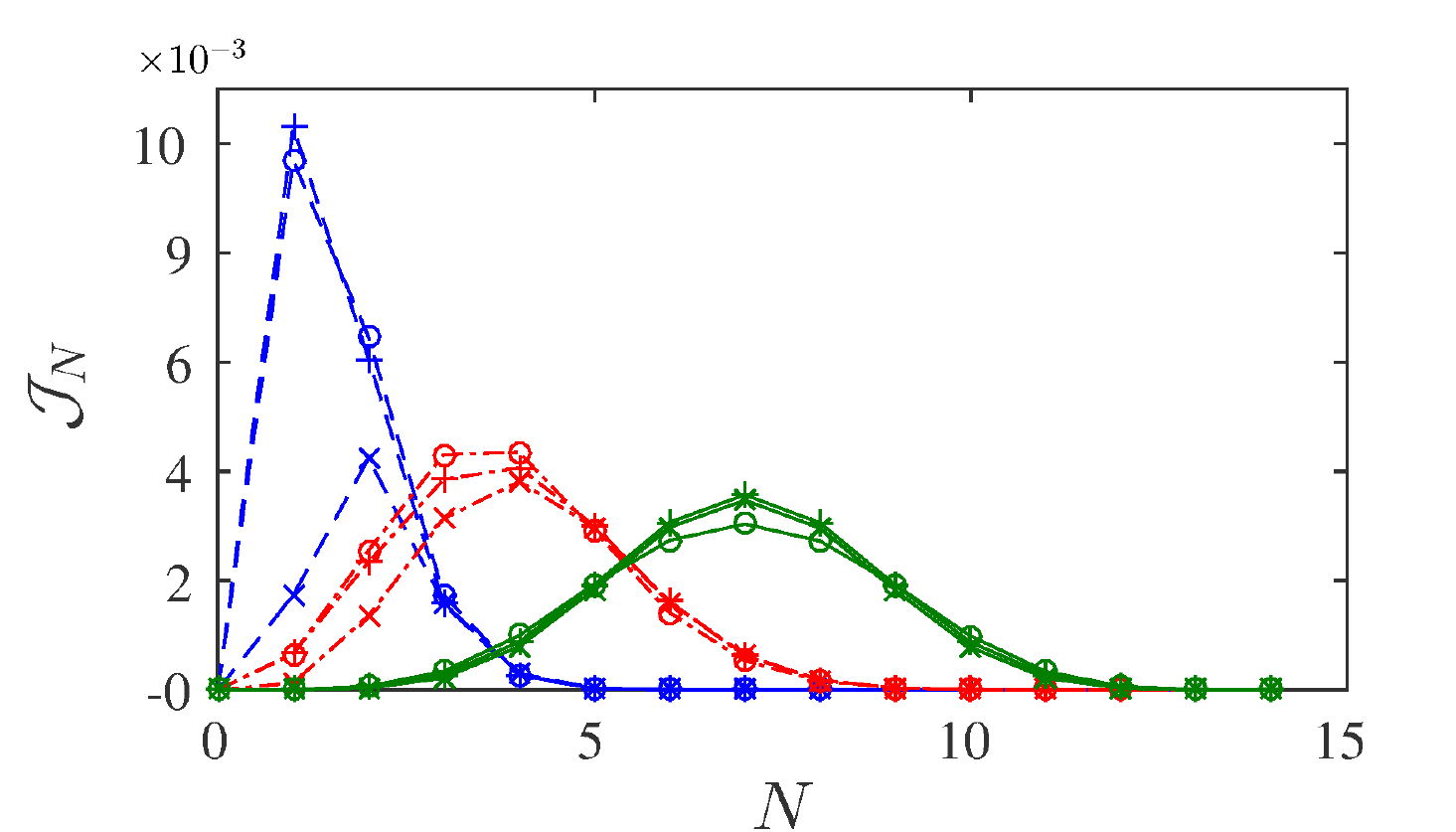}
\caption{(color online) Particle current per number sector $\cur_N$ for average filling $\bar{n}_{av}=0.05$ (blue dashed lines), $\bar{n}_{av}=0.25$ (red dot-dashed lines) and $\bar{n}_{av}=0.5$ (green continuous lines). For each $\bar{n}_{av}$ we show the current for different gauge fields: $\phi=0$ ($\circ$), $\phi=\pi/2$ ($+$) and $\phi=\pi$ ($\times$). Other parameters are $L=7$ and $K=J$.} \label{fig:fig3}       
\end{figure}

The exact ansatz (\ref{eq:rhoss}) also allows us to gain a deeper understanding of the working of the system. In fact it is now possible to compute the current for each number sector and thus realize which sectors contribute most to the current \cite{offdiag}. We thus fix $\Delta\bar{n}$ as in Fig.\ref{fig:fig2}(a) and we compute the current in each number sector $N$, $\mathcal{J}_N$, for various values of $\bar{n}_1$. more precisely $\mathcal{J}_N=\sum_p\lan\im Je^{\im(-1)^{p+1}\phi/2}\aop_{j,p}^{\dagger}\aop^{}_{j+1,p} +{\rm H.c}\ran_N /\hbar$ and $\lan\dots\ran_N$ means expectation over $\rhop_{ss}^N$. In Fig.\ref{fig:fig3} we show $\mathcal{J}_N$ as a function of the sector's particle number $N$, for different values of the gauge field $\phi$. We observe that for low fillings the current is mostly due to the sector with $1$ particle and also that the total current strongly varies as the gauge field changes. In particular for $\phi=\pi$ (blue dashed line with $\times$), the current is significantly lower than for $\phi=0,\pi/2$ (blue dashed lines with respectively $\circ$ and $+$ ). For larger $\bar{n}_{av}$ the particle number sectors which contributes most to the current are those of larger particle number and, for them, the current is much less affected by a change in the gauge field. 

\begin{figure}
\includegraphics[width=\linewidth]{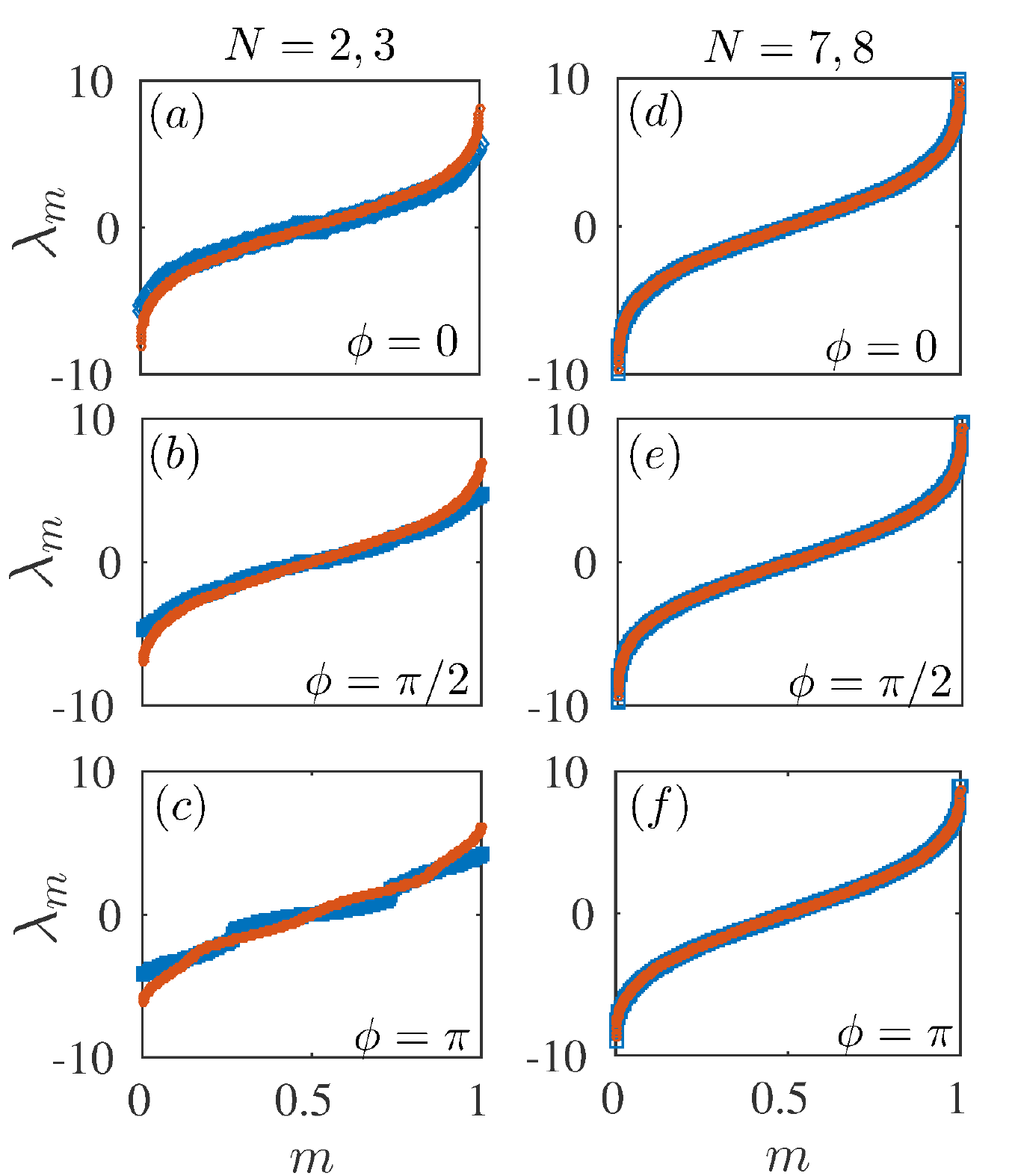}
\caption{(color online): Energy spectra for different particle filling and gauge fields $\phi$. For (a-c) the ladder has $10$ rungs and total particle number $2$ (blue squares) or $3$ (red circles), i.e. close to $1/10$ filling, while for (d-f) the ladder has $7$ rungs and total particle number $7$ (blue squares) or $8$ (red circles), i.e. close to half-filling. For panels (a,d) $\phi=0$, (b,e) $\phi=\pi/2$ and (c,f) $\phi=\pi$. For all these panels $K=J$.     } \label{fig:fig4}  
\end{figure}

The repartition of the density matrix in different number sectors, as in Eq.(\ref{eq:rhoss}), can give even further insight. Since the different number sectors are only coupled by the dissipator, and since the effectiveness of the coupling is strongly dependent on the spectrum in each number sector, by analyzing the spectrum of the Hamiltonian in each sector we can foresee whether the change of the phase $\phi$ would significantly affect the steady state and hence the current. In the left panels of Fig.\ref{fig:fig4} [panels (a, b, c)] we show the energy spectrum for total particle numbers $2$ and $3$ for a ladder of $10$ rungs. The spectrum changes significantly, especially for $N=2$, hence we expect a great change in the steady state and in the current as $\phi$ changes. In the right panels instead, (d, e, f), we show the spectrum for a ladder of $7$ rungs and either $7$ or $8$ hardcore bosons, corresponding to half-filling and half-filling plus one atom, two sector numbers which would also be directly coupled by the dissipator. In this case, in contrast to the low-filling case represented in the left panels, the energy spectrum does not change so significantly and the curves of the spectrum are always close to each other \cite{gapspectrum}. This is why, at larger fillings a much lower variation of the current as a function of the gauge field $\phi$ is expected, which justifies the results in Figs.\ref{fig:fig2} and \ref{fig:fig3}. It should be stressed that for non-equilibrium steady states, unlike in (zero temperature) quantum phase transitions, it is in general important to consider the full spectrum and not just the low energy part.

\begin{figure}
\includegraphics[width=\columnwidth]{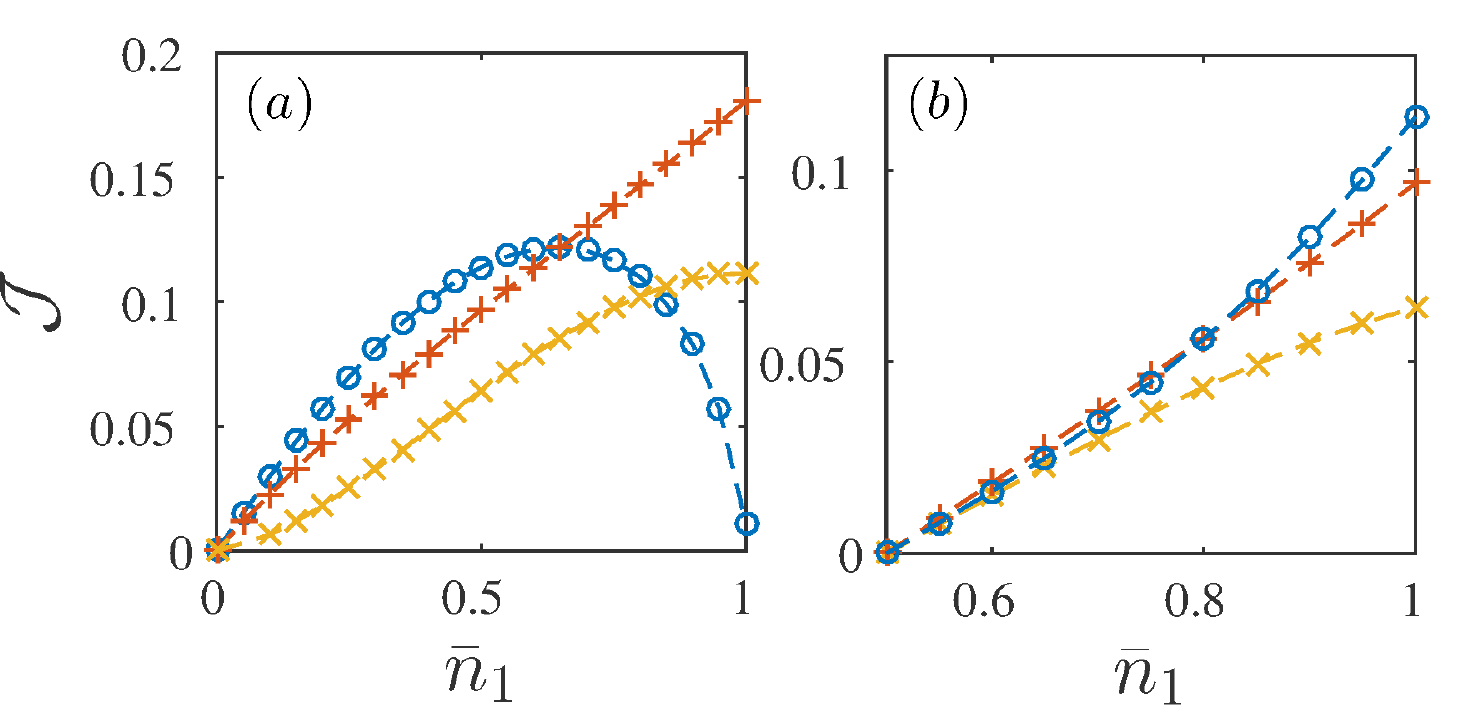}
\caption{(color online) Current vs $\bar{n}_1$ for phases $\phi=0$ ($\circ$), $\phi=\pi/2$ ($+$) and $\phi=\pi$ ($\times$). In panel (a) $\bar{n}_L=0$ while in panel (b) $\bar{n}_L=0.5$. Other parameters are $L=7$ and $K=1.5J$. 
} \label{fig:fig5} 
\end{figure}

{\it Negative differential conductance}: Because of the marked different dependence of the current as a function of the phase $\phi$ while the filling is varied, the conductance presents a nonlinear response to the dissipative boundary driving. This is shown in Fig.\ref{fig:fig5} where the current is depicted as a function of $\bar{n}_1$ for different values of $\bar{n}_L$. In particular we show in panels (a) and (b) respectively the cases for $\bar{n}_L=0$ and $\bar{n}_L=0.5$. In both panels we observe a strong nonlinear response with $\Delta \bar{n}$ which is very different for different values of the phase $\phi$. It should be noted that for free fermions, which can sometimes be used to describe the behavior of hardcore bosons, the response would be linear.  

Interestingly, in panel (a) we observe a strong signature of negative differential conductance for $\phi=0$. Increasing the gauge field the response becomes roughly linear, see the red $(+)$ for $\phi=\pi/2$, and then, at $\phi=\pi$, the response is superlinear, i.e. the current increases more than linearly when $\Delta \bar{n}$ increases.   
In panel (b), for $\bar{n}_L=0.5$, the superlinear behavior is even clearer, and this time it occurs for $\phi=0$, the case for which, at lower $\bar{n}_L$, negative differential conductance occurred.             

This behavior could have been anticipated from the results in Fig.\ref{fig:fig2}. In fact it is there shown that for small values of the phase $\phi$, the current is larger for lower fillings while at large $\phi$ the current is in general lower at lower fillings.

{\it Conclusions}: The interplay between gauge fields and dissipation can induce non-equilibrium phase transitions and markedly change the properties of a system. Also the interplay between a gauge field and interactions can induce quantum phase transitions. In this work we study the interplay of dissipation, gauge field and interactions. In particular, we have shown how interactions can strongly alter the ability to tune the transport properties of a system using, for example, a gauge field. Previous works had shown that a gauge field can be used to strongly vary the current flowing through two coupled chains, and phase transitions could emerge. Here we have shown that because of strong interactions, as the filling increases, the sensibility of the system to the gauge field is significantly reduced. Due to the interplay between the gauge field and the filling, the conductance has a strong non-linear behavior as a function of the system parameters, resulting also in negative differential conductance. Our calculations are exact and greatly simplified thanks to the use of quantum number conservation for the steady state density matrix (a method which can be readily implemented in many set-ups).         
In the future it would be important to extend the current work to include the role of finite interactions and longer chains. It would be particularly interesting to understand the fate of the non-equilibrium phase transitions predicted for free bosons as the interaction is smoothly changed from $0$ to a finite large value. A different nature of the particles (e.g. interacting fermions), or of the baths (e.g. non-Markovian thermal baths) should also be analyzed to understand deeply the transport properties of dissipatively boundary driven manybody quantum systems.

{\it Acknowledgments}: D.P. ackowledges support from Ministry of Education of Singapore AcRF MOE Tier-II (project MOE2016-T2-1-065, WBS R-144-000-350-112).



\begin{thebibliography}{10}
\bibitem{gamma} To ensure that the filling at the edges is exactly $\bar{n}_j$ we set $\hbar\Gamma=10 J$.
\bibitem{unique} For the parameters studied the steady state is unique.
\bibitem{offdiag} Note that the current operator conserves the number of particles, hence only the blocks with the same total number of particles in the bra and in the ket contribute to the current.
\end{thebibliography}
\end{document}